\documentclass[a4paper]{article}
\usepackage[numbers,square]{natbib}
\usepackage{INTERSPEECH2021}
\usepackage{amsmath,graphicx}
\usepackage{url}
\usepackage{tabularx} 
\usepackage{colortbl,color}
\usepackage{multirow}
\usepackage{tabularx}
\newcommand{\MJ}{\color{black}}
\newcommand{\MG}{\color{black}}
\newcommand{\MJJ}{\color{black}}
\newcommand{\GG}{\color{black}}
\newcommand{\IS}{\color{black}}
\newcommand{\fin}{\color{black}}

\title{Extracting Different Levels of Speech Information from EEG Using an LSTM-Based Model}
\name{Mohammad Jalilpour Monesi$^{1,2}$, Bernd Accou$^{1,2}$, Tom Francart$^2$, Hugo Van Hamme$^1$}
\address{$^1$KU Leuven, PSI, Dept. of Electrical engineering (ESAT), Leuven, Belgium\\
	$^2$KU Leuven, ExpORL, Dept. Neurosciences, Leuven, Belgium}
\email{mohammad.jalilpourmonesi@esat.kuleuven.be, hugo.vanhamme@esat.kuleuven.be}

\begin{document}

\maketitle
\begin{abstract}
  Decoding the speech signal that a person is listening to from the human brain via electroencephalography (EEG) can help us understand how our auditory system works.
  Linear models have been used to reconstruct the EEG from speech or vice versa. Recently, Artificial Neural Networks (ANNs) such as Convolutional Neural Network (CNN) and Long Short-Term Memory (LSTM) based architectures have outperformed linear models in modeling the relation between EEG and speech. Before attempting to use these models in real-world applications such as hearing tests or (second) language comprehension assessment we need to know what level of speech information is being utilized by these models. In this study, we aim to analyze the performance of an LSTM-based model using different levels of speech features. The task of the model is to determine which of two given speech segments is matched with the recorded EEG. We used low- and high-level speech features including: envelope, mel spectrogram, voice activity, phoneme identity, and word embedding. 
  Our results suggest that the model exploits information about silences, intensity, and \MJ broad phonetic classes \fin from the EEG. 
  Furthermore, the mel spectrogram, which contains all this information, yields the highest accuracy (84\%) among all the features.
\end{abstract}

\section{Introduction}


Understanding how natural running speech is processed in human auditory system has attracted a lot of attention in recent years. In the most common approach, natural running speech is presented to a listener and EEG signals are recorded simultaneously. A linear model is used to either reconstruct a speech representation from the recorded EEG (backward model) or to reconstruct the EEG from the speech (forward model). Then the correlation between the original and the reconstructed signal is used as a measure of neural speech tracking \cite[e.g.][]{diliberto_low-frequency_2015, crosse_multivariate_2016, vanthornhout_speech_2018, lesenfants_predicting_2019}. It has been shown that this approach can be used as an objective measure of how well speech is understood by a listener \cite{vanthornhout_speech_2018, lesenfants_predicting_2019, di_liberto_neural_2021}. Real-time and accurate speech decoding from the brain has other potential applications such as Brain-Computer Interfaces (BCIs).


The speech envelope has been the most common choice as a representation of speech in this context \cite[e.g.][]{ding_emergence_2012, crosse_multivariate_2016, vanthornhout_speech_2018}. Nevertheless, low-level acoustic features such as the spectrogram \cite{diliberto_low-frequency_2015, lesenfants_data-driven_2019}, or higher-level representations such as phonemes \cite{diliberto_low-frequency_2015, khalighinejad_dynamic_2017}, 
phonetic features \cite{teoh_attention_2020, lesenfants_data-driven_2019}, phonotactics \cite{di_liberto_low-frequency_2019, di_liberto_neural_2021}, and semantics have also been used in linear backward/forward models. In \cite{diliberto_low-frequency_2015, di_liberto_indexing_2017, lesenfants_data-driven_2019}\MG , the \fin authors have shown that including both low- and high-level speech features improves predicting brain responses to speech. 
Lesenfants et al. \cite{lesenfants_predicting_2019} showed that combining phonetic features with \MG the \fin spectrogram improves speech reception threshold (SRT) prediction from EEG.


In the aforementioned studies linear models are used in a regression problem. Even though linear models are easy to interpret and implement, they are inherently not able to model complex and non-linear auditory processing in the brain. To address this, Artificial Neural Networks (ANNs) have been explored as an alternative approach. ANNs have been used in the context of Auditory Attention Decoding (AAD), in which the task is to determine which of two concurrent speakers a listener attends to using EEG signals. Simple feedforward neural networks \cite{de_taillez_machine_2017} as well as Convolutional Neural Networks (CNNs) \cite{ciccarelli_comparison_2019} have been used in AAD. In \cite{monesi_lstm_2020} we presented an LSTM-based model to relate EEG responses to speech stimuli through a match/mismatch classification problem. The match/mismatch classification task is defined here as which of two given speech stimuli caused a given EEG recording. We showed that the proposed model works better than the linear model and generalizes well to unseen subjects. 

Our goal in this study is to understand what speech information \MJ we can infer from the recorded EEG using the LSTM-based model \cite{monesi_lstm_2020}. \fin
This will enable us to design speech decoding models for applications such as objective hearing tests and speech intelligibility \MG measurement\fin . \MJ Therefore, we analyze \MG the \fin performance of the LSTM-based model in the match/mismatch classification task using different levels of speech information. More specifically, we do a series of experiments where each uses a subset of speech information from the previous one \MG and \fin observe the difference in classification accuracy between \MG these \fin experiment. This enables us to see which speech information contributes to match/mismatch classification accuracy.\fin 
\section{Methodology}

In this section, we will first explain our data collection and preprocessing procedure. Then, we will introduce our classification task in detail.
Finally, we present how we adjust the LSTM-based model \cite{monesi_lstm_2020} for new speech features such as the mel spectrogram.

\subsection{Data collection and preprocessing}

We recorded EEG signals from 86 normal-hearing native-Flemish-speaking subjects while they listened to natural running speech in the form of stories. Subjects were screened for normal hearing with pure tone audiometry and the Flemish matrix-test \cite{luts_development_2014}.
Throughout our recordings, stories were chosen from a set of 10 stories each lasting approximately 14 minutes and 30 seconds. 43 subjects listened to 8 stories, 38 listened to 7 stories, 4 listened to 6 stories, and 1 subject listened to \MG only \fin 2 stories.

We used \MG the \fin APEX 4 software \cite{francart_apex_2008}, developed at ExpORL, to present the stimuli (stories). The stimuli were presented binaurally at 62 dBA using  Etymotic ER-3A insert phones. After each story, we asked subjects a question related to the story to make sure they paid attention. EEG signals were recorded using a 64 channel Active-two system from Biosemi at 8 kHz sampling rate. EEG recordings were measured inside an electromagnetically shielded and soundproofed cabin.

A multi-channel Wiener filter \cite{somers_generic_2018} was used to remove artifacts from the EEG recordings. Then, all EEG channels were re-referenced to common-average. Next, we bandpass filtered the EEG signals between 0.5-32 Hz \IS using a Chebyshev2 filter with 80dB stopband attenuation \fin and used Matlab's resample function to downsample them to 64 Hz. Matlab uses anti-aliasing low-pass filter before resampling. As a result the true bandwith is cut to 0.5-30 Hz. \IS Note that many studies \cite{ding_adaptive_2013, vanthornhout_speech_2018} have shown that Delta band (0.5-4 Hz) performs better than higher frequencies when relating speech to EEG. However, we included Theta, Alpha, and Beta frequency bands to provide more information to the model\fin . As a last step, we applied mean-variance normalization for each single EEG recording, i.e., \MG a \fin recording of around \IS 14 minutes and 30 seconds \fin in response to one story.

Speech features: We used the following representations of speech:

1. Envelope: we used the 'powerlaw subbands' method \cite{biesmans_auditory-inspired_2017} to extract the amplitude envelope, resulting in a one-dimensional feature.

2. Mel spectrogram: the mel scale was applied to Short Time Fourier Transform (STFT) in the range of 50 to 5000 Hz. We used 28 frequency bands. 

3. Voice Activity Detection (VAD): This one-dimensional binary feature is \MG one \fin when the energy after pre-emphasis in a 15~ms frame is above the 75\% percentile of values observed in the story, else it is zero.

4. Phoneme: Using forced alignment we segmented each word into a sequence of phonemes based on the International Phonetic Alphabet (IPA). Then for each of the 40 phonemes we converted the symbol to a one-hot vector of dimension 40.


5. Word embedding: 
We used pre-trained word embeddings of dimension 300 for Dutch which were trained using FastText \cite{grave2018learning}.

The envelope and mel spectrogram were bandpass filtered between 0.5 and 32 Hz \IS using a Chebyshev2 filter with 80dB stopband attenuation \fin. All the speech features were downsampled to 64 Hz to be synchronous with the EEG.

\subsection{Classification task}

We split each recording into training, validation, and test sets. The validation and test sets contain 10\% each, which were taken from the middle of the recording, and the remaining 80\% were taken as the training set.

\begin{figure}[htb]
	
	\begin{minipage}[b]{1\linewidth}
		\centering
		\centerline{\includegraphics[width=\textwidth]{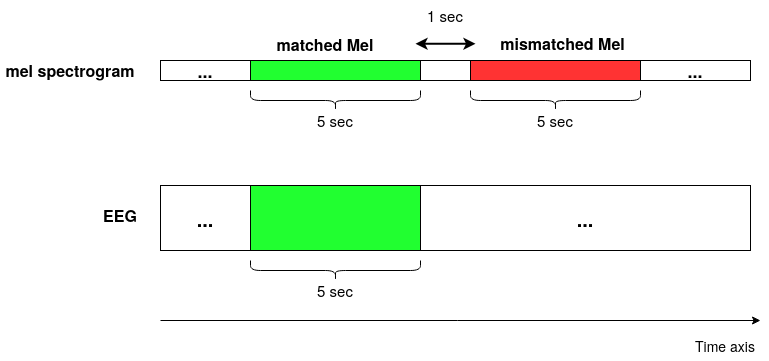}}
	\end{minipage}

	\caption{Extracting match and mismatch mel spectrograms with respect to an EEG segment of 5 seconds (decision window).}
	\label{fig:match_mismatch}
\end{figure}

As the classification task is defined exactly the same for each of the speech features, we will only consider mel spectrogram (mel) here.
We cut both EEG and mel into 5 seconds decision windows with 90\% overlap. For each EEG segment of 5 seconds, we take the corresponding 5 seconds from the mel sequence as `matched mel'. Next, we take a segment of mel that starts one second after the end of the matched mel as the `mismatched mel' (see figure \ref{fig:match_mismatch}). Then, given (EEG, matched mel, mismatched mel) the task of the model is to determine which of the two mel spectrograms matches the EEG. Our outcome measure is the classification accuracy of this task. 

\subsection{Model}

\begin{figure*}[htb!]
	
	\centering
	\centerline{\includegraphics[width=\textwidth, height=4cm]{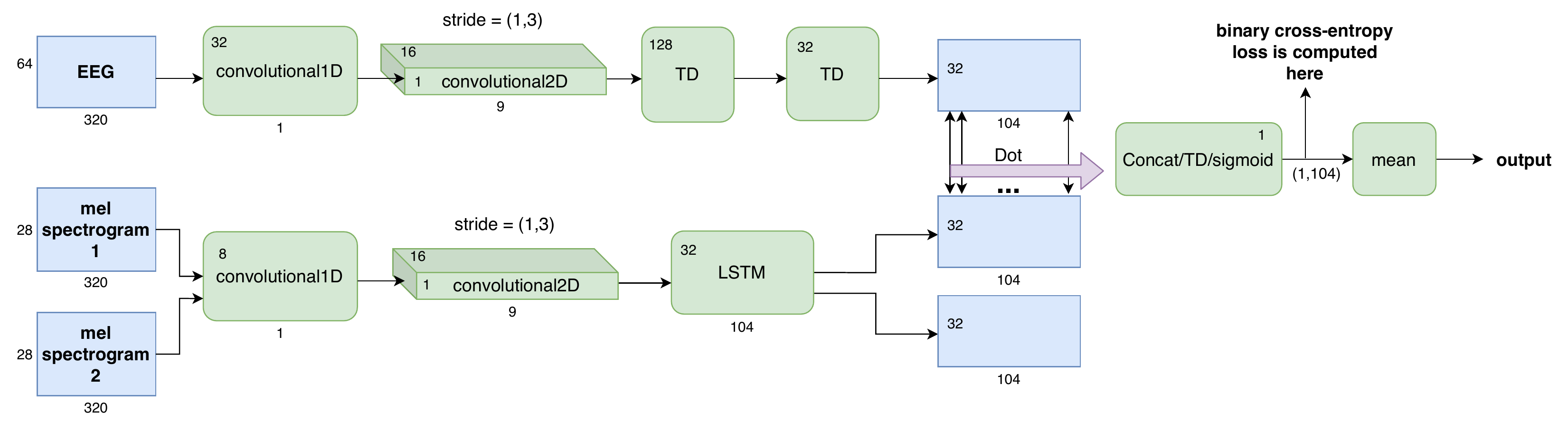}}
	
	%
	\caption{The LSTM-based model \cite{monesi_lstm_2020} for match/mismatch classification. TD refers to time distributed which applies a dense layer to every temporal slice of the input. Dot is a layer that applies dot product (cosine similarity) between EEG representation and speech representation for each time step.}
	\label{fig:LSTM model}
\end{figure*}

In this study, we used the LSTM-based architecture described by \cite{monesi_lstm_2020},  adjusted for different speech features. In figure \ref{fig:LSTM model} the LSTM-based model is shown with the mel spectrogram as a speech feature. The core ideas behind this architecture were (1) to use an LSTM layer in the speech path to model the brain response to the stimulus, including its delay, and (2) to  map short segments of EEG and speech features to a common embedding space where we expect that matched speech (mel spectrogram) and EEG have similar representations while mismatched speech and EEG have dissimilar representations. For more details about the model and hyper-parameter tuning please see \cite{monesi_lstm_2020}.

Here, compared to the original LSTM-based model, we changed the first layer in the EEG path (first 4 layers in figure \ref{fig:LSTM model} that apply to EEG) from a 2D convolution to a 1D convolution, for interpretation reasons. For \MJ some \fin speech features we slightly adjusted the network: the EEG path remained the same but we added  or removed some layers depending on the speech feature. In case of the mel spectrogram \MJ and phoneme-based features with feature dimension of at least 2, the architecture is shown in figure \ref{fig:LSTM model}. For one-dimensional features such as Envelope and VAD, we do not have the first 1D convolutional layer in the speech path \fin. For the word embedding feature, we replaced the 1D and 2D convolutional layers in the speech path with a max pooling layer with a stride of 3. 
For all the speech features, layers in the speech path are shared between the two speech inputs. We have provided the code for our models in
\MJ https://github.com/jalilpour-m/match-mismatch-speech-features. \fin

\section{Results}
\label{sec:results}

For each speech feature, we trained the model in a subject independent fashion, i.e., we only trained one general model using all the training data of all the subjects. Also, when we concatenated two (or more) features, they are treated separately by the the model up to the input of the LSTM layer, where for each time step we concatenate them in the feature space. We assume the LSTM is capable of integrating both information sources to the benefit of the match/mismatch classification accuracy.

\subsection{VAD, envelope, and mel spectrogram}

\begin{figure}[htb]
	
	\begin{minipage}[b]{1\linewidth}
		\centering
		\centerline{\includegraphics[width=\textwidth]{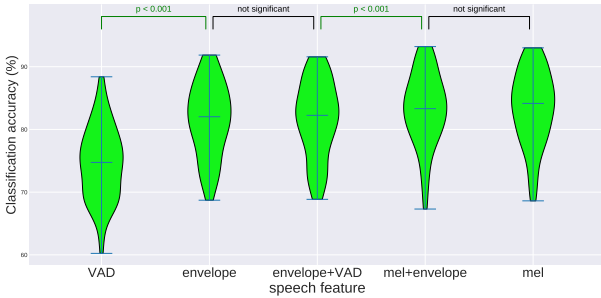}}
	\end{minipage}

	\caption{Classification accuracy of speech features voice activity detection (VAD), envelope, and mel spectrogram (referred to as mel). Violin plots are shown over 86 subjects. Decision window was set to 5 seconds.
	}
	\label{fig:vad_env}
\end{figure}

We wanted to know what level of speech information - reflected in EEG and speech features - the model exploits for the task. Therefore, we investigated how different speech features performed in our match/mismatch classification task. Previously we had evaluated the LSTM-based model working with the envelope as a speech feature. Here, we use the mel spectrogram which is a richer feature than the envelope (which can be computed from mel spectrogram), potentially leading to higher performance. In turn, from the envelope, we can derive the VAD feature.
Figure \ref{fig:vad_env} shows that the model's median accuracy is 75\% and 82\% for VAD and envelope respectively. \MJ These \fin results suggest that \MG a substantial share \fin of the accuracy comes from extracting silences from the EEG. Furthermore, when we concatenate VAD with envelope we do not gain any accuracy since VAD could be extracted from envelope. Moreover, concatenating mel spectrogram with envelope performs the same as mel spectrogram, as the mel spectrogram includes all the information available in the envelope. More specifically, there is a slight increase in performance when we use mel spectrogram instead of envelope, which is statistically significant ($z=-6.48$, $p<0.001$). We used Wilcoxon signed-rank test with normal approximation for all our statistical tests.

\subsection{Phoneme and word embedding}
\label{ssec:concat}

We also used a 40-dimensional phoneme feature as this has been shown to improve performance in linear models \cite{lesenfants_data-driven_2019, lesenfants_predicting_2019}. We also created three simplified features from the phoneme feature where each feature contains less information compared to the previous one. First, we grouped phonemes into 5 categories: short vowels, long vowels, plosives, fricatives, nasals and approximants. \MJJ We concatenated these 5 categories with silence and created a six-dimensional feature called broad phonetic classes (BPC) in this study. Second, we grouped vowels and consonants and together with silences creating a three-dimensional feature called vowel/consonant. Third, we combined vowels and consonants in the vowel/consonant feature to one category and with silences created a two-dimensional feature called anyPhoneme. AnyPhoneme is a one-hot vector which determines at each time step whether it is silence or any phoneme. \fin The idea behind creating these simpler phoneme representations is to investigate what phonetic \MG detail is used \fin by the network. \MJJ However, when we combine different phonemes into one category we lose phone onsets. This is especially the case for anyPhoneme feature \GG which only retains onsets after pauses, which are only allowed between words by the forced alignment model. \fin In the case of vowel/consonant, whenever there are two or more consecutive consonants we will not have onset information of the consecutive consonants. We will only have onsets when altering from vowel to consonant and vice versa. Similarly, we might lose some onsets in the BPC feature which will happen rarely. In order to ensure that performance of the model is not affected by losing onset information we also created all three features from phoneme onsets. The versions that we kept the phone onsets are named anyPhoneme\_onset, vowel/consonant\_onset, and BPC\_onset in this study.

\begin{figure}[htb]
	
	\begin{minipage}[b]{1\linewidth}
		\centering
		\centerline{\includegraphics[width=\textwidth]{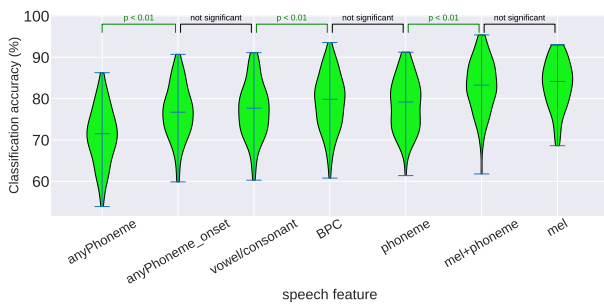}}
	\end{minipage}

	\caption{classification accuracy of the model for 5 phoneme related features and mel spectrogram. BPC stands for broad phonetic classes. Violin plots are shown over 86 subjects and the decision window is 5 seconds.}
	\label{fig:phoneme}
\end{figure}

Performance of the model using phoneme related features is shown in figure \ref{fig:phoneme}. Note that we do not report accuracy for vowel/consonant\_onset and BPC\_onset since they had the same performance as vowel/consonant and BPC respectively. However, \GG AnyPhoneme\_onset outperforms anyPhoneme ($z=-8.01$, $p<0.001$) suggesting that the model uses phone onset information. Next, we observe that vowel/consonant (or vowel/consonant\_onset) has the same performance as anyPhoneme\_onset. This suggests that the distinction between vowels and consonants is not informative enough. Nevertheless, it can use the phone onsets that occur in vowel/consonant to increase the accuracy compared to anyPhoneme feature. \fin BPC outperforms vowel/consonant ($z=-5.37$, $p<0.001$). This suggests that the model is capable of using some broad phonetic class information from the EEG. Finally, we observed no difference in performance when we used phoneme feature compared to the BPC.  We argue that the model is unable to use more detailed phoneme information beyond BPC. Furthermore, concatenating mel with phoneme yields the same performance as that of the mel. This suggests that the model can extract phoneme (or BPC) level information from the mel spectrogram.\fin

\begin{figure}[htb]
	
	\begin{minipage}[b]{1\linewidth}
		\centering
		\centerline{\includegraphics[width=\textwidth]{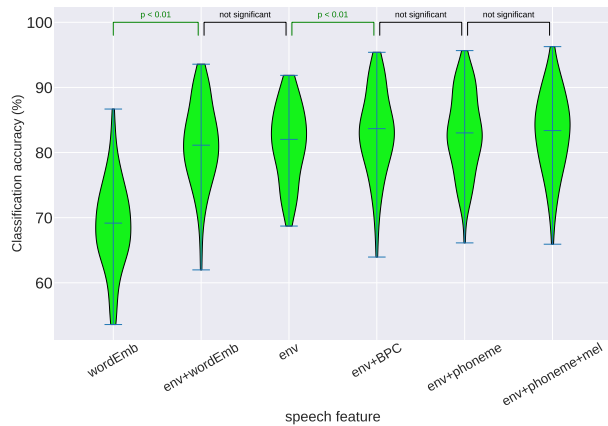}}
	\end{minipage}

	\caption{Classification accuracy of the model for word embedding (wordEmb), envelope, envelope+BPC, envelope+phoneme, and envelope+phoneme+mel spectrogram. In the figure, env, BPC, and mel refer to envelope, broad phonetic classes, and mel spectrogram respectively.}
	\label{fig:wordEmb}
\end{figure}

To test whether the model can extract any contextual information from speech, we used a word embedding feature. Figure \ref{fig:wordEmb} shows that the performance of the word embedding, referred \MJ to as \fin wordEmb, is around only 70\% median accuracy. Also, when we concatenate this feature with envelope, we observe the same performance as for \MJ that of \fin the envelope. In addition, the anyPhoneme feature which is basically a simplified binary version of wordEmb performs the same as wordEmb. These two observations suggest that the model does not use any word level information from the wordEmb feature.

\MG Furthermore, we concatenated the envelope and the BPC (or phoneme) feature expecting that the performance will be close to that of the mel spectrogram. Figure \ref{fig:wordEmb} shows that concatenating the envelope with the BPC (env+BPC) or phoneme (env+phoneme) indeed increases the classification accuracy to almost that of the mel spectrogram. In addition, classification accuracy of combined envelope, phoneme, and mel spectrogram features is the same as for the mel spectrogram. These results are plausible since one could argue that the mel spectrogram contains intensity as well as BPC level information. If we average all the frequency bands of the mel spectrogram we obtain envelope which contains intensity (energy) of the speech. It has been shown a long time ago \cite{martens_broad_1991} that a simple multilayer perceptron with one layer and a few nodes can do BPC classification using mel spectrogram input. Note that the LSTM layer in our model is capable of behaving as a simple multilayer perceptron by always forgetting the past. In short, these observations suggest that within the significance of this experiment, the mel-model is exploiting intensity information as well as some information related to broad phonetic classes. \fin


%

\section{Conclusion}
\label{sec:con}

Our objective was to investigate what speech information we can decode from the recorded EEG. As a result, we trained the LSTM-based model \cite{monesi_lstm_2020} in the match/mismatch classification task using different levels of speech features. We used a broad range of speech features from a very low-level acoustic features such as VAD to higher-level features such as word embedding. \MJ In one set of experiments, we started with VAD because we wanted to test how much of the classification accuracy comes from just tracking silences in the speech. Our results show that indeed using only VAD the model can reach a classification accuracy of 75\%. Then, we used envelope which contains intensity information over time. We observed that the classification accuracy increased to 82\%. This suggest that the LSTM-based model is capable of extracting information about intensity of the speech. Finally, using mel spectrogram, which is a richer signal than the envelope, the model reached classification accuracy of 84\%. The observed results are expected because envelope contains all the information present in VAD and in turn mel contains all the information present in the envelope.\fin

\MJ In another set of experiments, we investigated which phoneme level information is being used by the model. We created \MG three \fin simpler phoneme-based features from the 40 dimensional phoneme feature: broad phonetic classes (BPC), vowel/consonant, and anyPhoneme. \GG Our results show that phone onsets are important information for the performance of the model. More specifically, using the anyPhoneme feature we lose 5\% accuracy compared to the anyPhoneme\_onset feature. Considering phonetic detail, \fin our results suggest that the model is able to use information up to \MG the \fin BPC feature. In other words, using \MG the \fin BPC feature or more detailed phoneme feature results in the same performance. We also showed that if we train the model with envelope and BPC features combined we reach the same performance as with that of the mel spectrogram. We argue that at least traces of BPC can be found in \MG the \fin EEG since if we remove this information from the speech path the classification accuracy decreases.\fin

We also used a high level word embedding feature hoping that the model can extract some word level information from the EEG. We used this feature both alone and also in combination with the mel spectrogram. Unfortunately, our results suggest that the model is unable to extract word level information from EEG. In summary, it seems that the model is able to extract silences, energy intensity, and some \MJ broad phonetic class \fin level information from EEG. Extracting word level information from EEG might be possible using more complex ANN models and more training data.

\section{Acknowlegements}
The work is funded by KU Leuven Special Research Fund C24/18/099 (C2 project to Tom Francart and Hugo Van hamme) and the Flemish Government under "Onderzoeksprogramma AI Vlaanderen". This project has also received funding from the European Research Council (ERC) under the European Union’s Horizon 2020 research and innovation programme (grant agreement No 637424, ERC starting Grant to Tom Francart).

\bibliographystyle{IEEEtran}

\bibliography{refs}


\end{document}